\documentclass[aps, pra, showpacs, twocolumn]{revtex4}
\usepackage{graphics}
\usepackage{graphicx}
\usepackage{float}
\usepackage{amssymb}
\usepackage{amsmath}
\usepackage{bm}

\newcommand{\bra}[1]{\langle #1 \vert}
\newcommand{\ket}[1]{\vert #1 \rangle}

\begin{document}

\title{Effects of ground state hyperfine shifts in quantum computing with optically hole burnt materials}
\author{Karl Tordrup}
\author{Klaus M{\o}lmer}
\affiliation{Lundbeck Foundation Theoretical Center for Quantum System Research, Department of Physics and Astronomy, University of  Aarhus,
DK-8000 Aarhus C, Denmark}

\date{\today}

\begin{abstract}
We present an investigation of the effects of constant but random shifts of the ground hyperfine qubit states in the setting of quantum
computing with ion doped crystals. Complex  hyperbolic secant pulses can be used to transfer ions reliably to electronically excited states, and
a perturbative approach is used to analyse the effect of ground state hyperfine shifts. This analysis shows that the errors due to the hyperfine
shift are dynamically supressed during gate operation, a fact we attribute to the AC Stark shift. Furthermore we present an implementation of a
controlled phase gate which is resilient to the effects of the hyperfine shift. Decoherence and decay effects are included in simulations in
order to show that a demonstration of quantum gates is feasible over the relevant range of system parameters.
\end{abstract}

\pacs{03.67.Lx, 32.80.Qk, 03.65.Yz}

\maketitle

\section{Introduction}

\begin{figure}
\includegraphics[width=8cm]{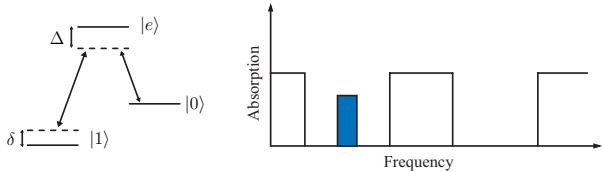}
\caption{Left: simplified level diagram of a single ion. Right: qubits are prepared as antiholes in a hole burnt structure. To ensure
appropriate interactions between qubits the system must first be properly initialised \cite{Kroll:optCommun}.} \label{fig:channel}
\end{figure}

\begin{figure}
\includegraphics[width=5cm]{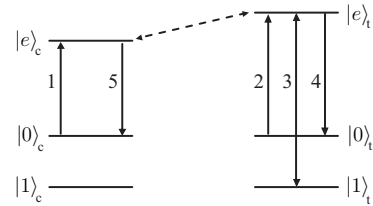}
\caption{Implementation of a controlled NOT gate. Pulses 2--4 implement the NOT gate on the target qubit. If the control qubit is initially in
the $\ket{0}$ state it is promoted by pulse 1 to the excited state and pulses 2--4 become off resonant. If, on the other hand, the control
starts out in the $\ket{1}$ state pulse 1 is off resonant and the NOT gate on the target ion proceeds.}\label{fig:CNot}
\end{figure}

An interesting class of proposals for quantum computing is provided by solid state devices with dopants as the active medium \cite{Kane},
\cite{Lukin}, \cite{longdell}, \cite{Kroll:optCommun}. In this article we focus on the Rare Earth Quantum Computing (REQC) system proposed by
Ohlsson \emph{et al.} \cite{Kroll:optCommun}. In this proposal qubit states are encoded in hyperfine levels of rare earth ions doped in an
inorganic crystal host. The qubit levels are coupled via an excited state accessible by optical excitation. Due to the large inhomogeneous
broadening of the excited state, qubits may  be selected in frequency space and identified by the magnitude of the inhomogeneous shift. In
practice qubits are prepared as narrow structures in a hole burnt structure within the inhomogeneous profile and readout is achieved using
absorption spectroscopy (see Fig.\ \ref{fig:channel}).
Each qubit consists of an ensemble of ions centered at a certain transition frequency. For this reason operation of the REQC system calls for
pulses that are robust to variations in detuning inherent in the proposal. Qubit interaction is mediated by the dipole--dipole coupling. The
rare earth ions within the crystal acquire a change in static dipole moment when in the excited state $\ket{e}$. This causes the excited state
of the surrounding ions to shift in energy, and provided the shift exceeds a certain threshold, resonant driving of the $\ket{i} \leftrightarrow
\ket{e}$ transition is blocked. This effect can be used to implement a fully entangling gate \cite{Kroll:optCommun} as illustrated in Fig.\
\ref{fig:CNot}

Previous theoretical works have dealt with the problem of transferring population reliably in spite of the finite width of the frequency
channels \cite{Roos} and the use of phase compensating pulses to eliminate the effects of unknown detuning dependent phases acquired during
pulses \cite{Wesenberg:bus}. In the laboratory qubit initialisation \cite{Kroll:init} and basic manipulations \cite{Kroll:transfer} have been
demonstrated with good fidelity and it appears that a fully entangling gate will soon be within reach.

In this paper we focus on an effect that has previously not been addressed in the REQC system. Apart from the excited state shift due to the
finite qubit channel width which is chosen in the initialisation procedure, there is also a broadening of the hyperfine qubit levels with a
width of about 30 kHz. The constant but random shift of the hyperfine levels within this range leads to dephasing between the $\ket{0}$ and
$\ket{1}$ states and is totally destructive to previously proposed gate implementations. In this paper we propose gate sequences that correct
for the ill effects of the hyperfine shift.

The detrimental effect of atomic decay is often neglected in theoretical investigations of REQC. Although this may be well justified for short
pulse sequences with negligible  integrated population of the excited state, it becomes important in longer error compensated schemes. We thus
present simulations which include a description of the main decoherence mechanisms.

 The paper is organised as follows. In Section
\ref{sec:sech} we give a theoretical description of the complex hyperbolic secant pulse and discuss the effects of the hyperfine shift. In
Section \ref{sec:qubitRot} we investigate the effects of the hyperfine shift on a previously reported method for performing arbitrary qubit
rotations on single qubits. In Section \ref{sec:gate} we propose a phase compensated implementation of a controlled phase gate which is robust
to the effects of the finite channel width as well as the hyperfine shift. Section \ref{sec:decoherence} deals with the main contributions to
decoherence and we perform a full simulation of our gate sequence taking these effects into account.

\section{The complex hyperbolic secant pulse}\label{sec:sech}

The complex hyperbolic secant pulse offers population inversion over a broad range of detunings \cite{Silver} and is therefore perfectly suited
for rare earth quantum computing. A detailed theoretical analysis given in \cite{Gouet} uses the Bloch sphere picture of two--level dynamics and
shows that the secant pulse, in fact, satisfies very well the adiabaticity criterion during an effective chirp through atomic resonance. The
Bloch sphere approach does not reveal the acquired global phase on the two coupled quantum states, which is important in the present quantum
information setting since in REQC we always address a subspace $\mathcal{H} = \{\ket{e},\ket{i}\}$ of the full three level system and hence a
global phase acquired during the $\ket{0} \leftrightarrow \ket{e}$ transitions results in a relative phase between the $\ket{0}$ and $\ket{1}$
qubit states.
 In the following we present a perturbative treatment which reveals the
global phases acquired during a Sech pulse.

The complex Rabi frequency of the Sech pulse is
\begin{equation}
\Omega_{\text{sech}}(t) = \Omega_0(\text{sech}(\beta t))^{1-i\mu} = \Omega_R(t)e^{i\varphi(t)},
\end{equation}
with the real function
\begin{equation}\label{eq:omegaR}
\Omega_R(t) = \Omega_0 \text{sech}(\beta t),
\end{equation}
 and the instantaneous frequency given by the time derivative of the phase
\begin{equation}
\dot{\varphi}(t) = \mu \beta \text{tanh}(\beta t).
\end{equation}
During the pulse the evolution is determined by the time dependent Hamiltonian. For a given ion with a transition frequency shifted by $\Delta$
with respect to the centre frequency of the channel we arrive at the Hamiltonian
\begin{equation}
H(t) = (\Delta + \dot{\varphi}(t))\vert e \rangle\langle e \vert + \tfrac{1}{2}\Omega_{R}(t)(\ket{i}\bra{e} + \text{hc}.),
\end{equation}
with $i = 0,1$ and $\hbar = 1$. Note that we are working in an accelerated frame rotating at $\dot{\varphi}(t)$. Diagonalising the instantaneous
Hamiltonian we find the time dependent energies
\begin{equation}\label{eq:energies}
E_{\pm}(t) = \frac{\Delta + \dot{\varphi}(t)\pm \sqrt{(\Delta + \dot{\varphi}(t))^2 + \Omega_{R}^2(t)}}{2},
\end{equation}
and the corresponding eigenstates
\begin{equation}\label{eq:eigStates}
\ket{\pm} = \frac{2E_{\pm}(t) \ket{e} +  \Omega_{R}(t) \ket{i}}{\sqrt{\Omega_{R}^2(t)+4E_{\pm}^2(t)}}.
\end{equation}
Expanding the Schr\"odinger equation in the time dependent eigenbasis we obtain the equations of motion
\begin{equation}
i\dot{c}_{\pm} = E_{\pm}c_{\pm} - \bra{\pm}\dot{\mp}\rangle c_{\mp}.
\end{equation}
In the adiabatic limit the coupling term $\bra{\pm}\dot{\mp}\rangle$ tends to zero and the action of the Sech pulse becomes an adiabatic
transfer. The equations of motion are then simply solved by integrating the time dependent energies
\begin{equation}\label{eq:zeroOrder}
c_{\pm}^{(0)}(t) = e^{-i\int_0^t E_{\pm}dt'}.
\end{equation}
We now include the effect of the diabatic correction as a perturbation. Switching to the interaction picture we have to first order
\begin{equation}\label{eq:Upert}
U_I = \mathbf{1} - i\int_0^t V_I dt'.
\end{equation}
The interaction picture potential in the basis $\{\ket{-},\ket{+}\}$ is
\begin{equation}\label{eq:firstOrder}
V_I =
\begin{pmatrix} 0 & -i\xi e^{i\int_0^t E_- - E_+dt'} \\
i\xi e^{i\int_0^t E_+ - E_-dt'} & 0 \end{pmatrix},
\end{equation}
with
\begin{equation}\label{eq:xi}
\xi = \frac{\dot{\Omega}_{R}(\Delta+\dot{\varphi})-\ddot{\varphi}\Omega_{R}}{2((\Delta + \dot{\varphi})^2+\Omega_{R}^2)}
\end{equation}

\begin{figure}
\includegraphics[width=7cm]{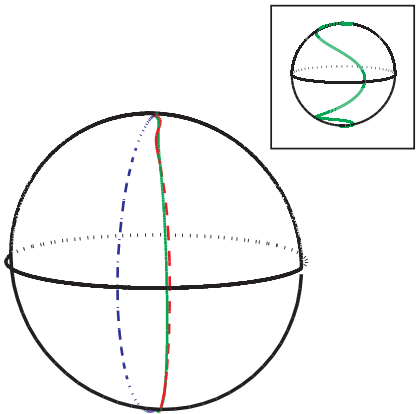}
\caption{(Color online) Bloch sphere trajectories during a Sech pulse with pulse parameters $\Omega_0 = 4$ MHz, $\mu = 3$, $\beta = 1.28$ MHz
and $T = 1.5$ $\mu$s. The North and South poles represent the states $\ket{e}$ and $\ket{i}$ respectively. The solid curve is the result of a
numerical integration with a standard ODE solver. The dot--dashed curve (left) represents the adiabatic $\ket{+}$ eigenstate of Eq.\
(\ref{eq:eigStates}) and the dashed curve represents the perturbative result of Eqs.\ (\ref{eq:Upert})--(\ref{eq:xi}). The inset shows the ODE
result in a frame rotating at constant frequency.}\label{fig:sechPert}
\end{figure}

In Fig.\ \ref{fig:sechPert} we plot the Bloch sphere trajectories during a Sech pulse computed to zeroth and first order and compare with a
numerical integration of the equations of motion. We observe good agreement between the numerical result and the first order perturbative result
so we can dispense with higher order corrections. Note that as long as we are restricting ourselves to the two dimensional subspace
$\mathcal{H}=\{\ket{e},\ket{i}\}$ the effects of the hyperfine shift can always be absorbed into the effect of the excited state shift. Only
when we regard the full three level system do the effects of the hyperfine shift materialise.

\section{Arbitrary qubit rotations}\label{sec:qubitRot}

The Sech pulse described in the previous section is only able to implement effective $\pi$--rotations. A method to perform arbitrary single
qubit rotations has been suggested in \cite{Roos}. By simultaneously applying two fields one can selectively address one of the two 'bar' states
defined by
\begin{subequations}\label{eq:barBasis}
\begin{equation}
\ket{\bar{0}} = \tfrac{1}{\sqrt{2}}(\ket{0} + e^{i\alpha}\ket{1})\\
\end{equation}
\begin{equation}
\ket{\bar{1}} = \tfrac{1}{\sqrt{2}}(\ket{0} - e^{i\alpha}\ket{1}).
\end{equation}
\end{subequations}
We selectively address the $\ket{\bar{0}}\leftrightarrow \ket{e}$ transition by simultaneously applying two fields with the $\ket{0}$--$\ket{e}$
and $\ket{1}$--$\ket{e}$ optical transition frequencies and with complex Rabi frequencies $\Omega_0(t) = \Omega_{\text{sech}}(t)$ and
$\Omega_1(t) = e^{i\varphi}\Omega_{\text{Sech}}(t)$ with $\varphi = \alpha$. In order to address the $\ket{\bar{1}}\leftrightarrow \ket{e}$
transition we choose the relative phase such that $\varphi = \alpha + \pi$.

Using Sech pulses we can selectively apply a phase to the $\ket{\bar{1}}$ state, i.e. effect the evolution
\begin{equation}\label{eq:arbPhase}
\bar{U} = \begin{pmatrix} 1 & 0 \\ 0 & e^{i\theta}\end{pmatrix},
\end{equation}
in the basis $\{\ket{\bar{0}},\ket{\bar{1}}\}$. The phase $e^{i \theta}$ is applied to the $\ket{\bar{1}}$ state by first applying a two--color
Sech--pulse to the $\ket{\bar{1}} \leftrightarrow \ket{e}$ transition. A second Sech--pulse is then applied with the phase shifted by $e^{i
\theta}$ relative to the first pulse returning the ions to the $\ket{\bar{1}}$ state with a geometric phase $e^{i \theta}$. During the two
pulses the $\ket{\bar{1}}$ state in addition to the geometric phase acquires a phase $\varphi(\Delta)$ dependent on the detuning from the qubit
channel center. Now, applying two Sech--pulses with a relative opposite phase to the $\ket{\bar{0}} \leftrightarrow \ket{e}$ transition the
$\ket{\bar{0}}$ state picks up the same detuning dependent phase $\varphi(\Delta)$, but no geometric phase, and $\varphi(\Delta)$ may be taken
as a global phase and thus be disregarded. The phases of the two fields $\Omega_0(t)$ and $\Omega_1(t)$ during this four--pulse sequence are
summarized in Table \ref{tab:arbPhase}.
\begin{table}
\caption{\label{tab:arbPhase}Phases $\varphi_0$ and $\varphi_1$ of the bichromatic Sech--pulse fields. The first two pulses apply a phase of
$\theta$ to the $\ket{\bar{1}}$ state. Pulses 3--4 compensate the detuning dependent phase acquired during the first two pulses.}
\begin{ruledtabular}
\begin{tabular}{ccccc}
            & 1 & 2 & 3 & 4 \\
            \hline
$\varphi_0$ & 0 & $\pi + \theta$ & 0 $\pi$\\
$\varphi_1$ & $\pi + \alpha$ & $\alpha + \theta$ & $\alpha$ & $\alpha + \pi$
\end{tabular}
\end{ruledtabular}
\end{table}

In the logical basis the evolution described by Eq.\ (\ref{eq:arbPhase}) is equivalent to
\begin{equation}
U =  e^{i\theta/2} \begin{pmatrix} \cos{\theta/2} & ie^{-i\varphi}\sin{\theta/2} \\ ie^{i\varphi}\sin{\theta/2} & \cos{\theta/2}
\end{pmatrix},
\end{equation}
which is an arbitrary rotation in the qubit space.

\subsection{Effects of inhomogeneous broadening}
The Sech pulse offers robustness against variations $\Delta$ in the resonant optical excitation frequencies. We now include in our formalism  a
term $\delta \ket{1}\bra{1}$ accounting for the inhomogeneous shift of the hyperfine ground levels. The Hamiltonian then becomes
\begin{eqnarray}
H(t) = && (\Delta + \dot{\varphi}(t))\ket{e}\bra{e} - \delta \ket{1}\bra{1} \nonumber \\
       && + \tfrac{1}{2}\Omega_0(t) \ket{e}\bra{0} + \text{hc.} \nonumber \\
       && + \tfrac{1}{2}\Omega_1(t) \ket{e}\bra{1} + \text{hc.}.
\end{eqnarray}
 Like $\Delta$, the shift $\delta$ varies for the different ions, and the goal is to implement gates which are also resilient to this
variation.
\begin{figure}
\includegraphics[width=8cm]{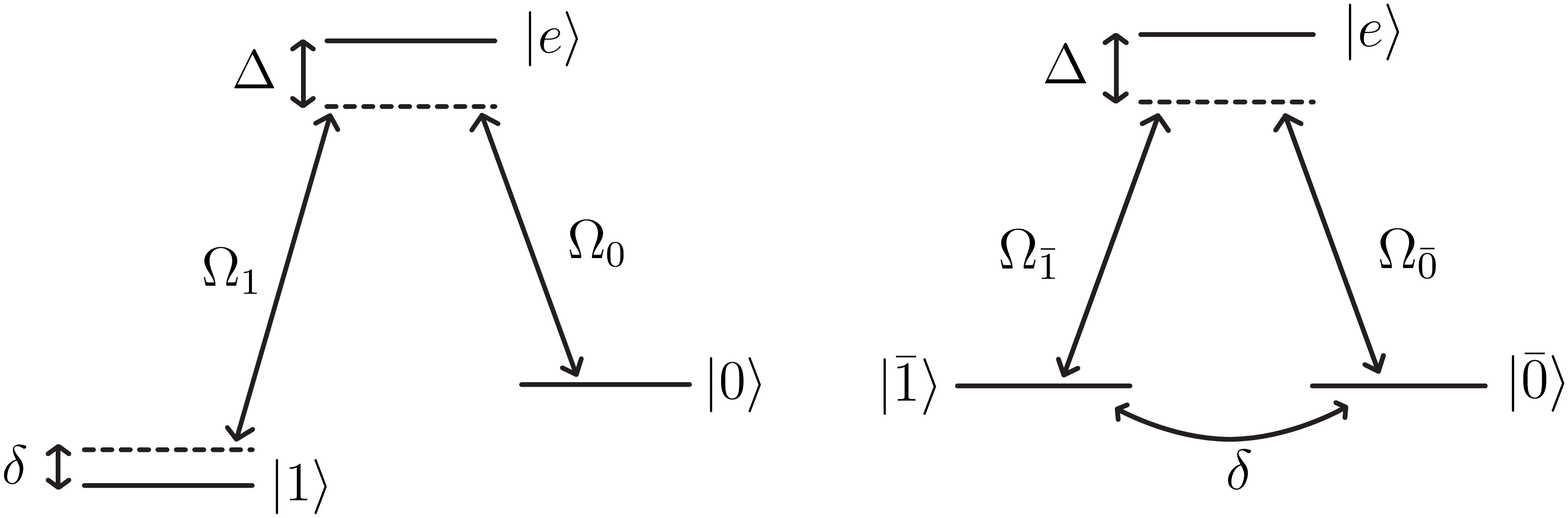}
\caption{In the logical basis inhomogeneous broadening of the qubit levels leads to dephasing(left). In the 'bar' basis the effect is a coupling
of the $\ket{\bar{0}}$ and $\ket{\bar{1}}$ states(right). }\label{fig:lambdaSysTwice}
\end{figure}
The Hamiltonian in the basis $\{\ket{e},\ket{\bar{0}},\ket{\bar{1}}\}$ takes the form
\begin{eqnarray}\label{eq:barHamiltonian}
H(t) = &&(\Delta+\dot{\varphi}(t))\ket{e}\bra{e} - \tfrac{\delta}{2}(\ket{\bar{0}}\bra{\bar{0}}+\ket{\bar{1}}\bra{\bar{1}})\nonumber\\
    && + \tfrac{1}{2\sqrt{2}}(\Omega_0(t) +e^{i\alpha} \Omega_1(t)) \ket{e}\bra{\bar{0}} + \text{hc.} \nonumber\\
    && + \tfrac{1}{2\sqrt{2}}(\Omega_0(t) -e^{i\alpha} \Omega_1(t))  \ket{e}\bra{\bar{1}} + \text{hc.} \nonumber\\
    && + \tfrac{\delta}{2}(\ket{\bar{0}}\bra{\bar{1}} + \ket{\bar{1}}\bra{\bar{0}}).
\end{eqnarray}
Since the inhomogeneous shift of the hyperfine ground states is of the order $\delta \sim 30$ kHz while the Rabi frequencies may be as large as
4 MHz we may treat the effect of hyperfine broadening perturbatively. We split the Hamiltonian of Eq.\ (\ref{eq:barHamiltonian}) into two parts
\begin{equation}
H(t) = H_0(t) + V,
\end{equation}
with
\begin{eqnarray}\label{eq:Ho}
H_0 = &&(\Delta+\dot{\varphi}(t))\ket{e}\bra{e} \nonumber\\
    && + \tfrac{1}{2\sqrt{2}}(\Omega_0(t) +e^{i\alpha} \Omega_1(t)) \ket{e}\bra{\bar{0}} + \text{hc.} \nonumber\\
    && + \tfrac{1}{2\sqrt{2}}(\Omega_0(t) -e^{i\alpha} \Omega_1(t))  \ket{e}\bra{\bar{1}} + \text{hc.}
\end{eqnarray}
and
\begin{equation}\label{eq:V}
V = \tfrac{\delta}{2}(-\ket{\bar{0}}\bra{\bar{0}}-\ket{\bar{1}}\bra{\bar{1}}+\ket{\bar{0}}\bra{\bar{1}}+\ket{\bar{1}}\bra{\bar{0}}).
\end{equation}
Since we only address one transition at a time the dynamics due to $H_0$ is restricted to the subspace $\mathcal{H} = \{\ket{e},\ket{\bar{i}}\}$
and is thus described by the methods developed in Section \ref{sec:sech} whereas V causes leakage of population between $\ket{\bar{0}}$ and
$\ket{\bar{1}}$. Resorting to Eqs.\ (\ref{eq:Upert}) and (\ref{eq:Ho})--(\ref{eq:V}) we find that when driving for example the $\ket{\bar{0}}
\leftrightarrow \ket{e}$ transition this leakage is described by
\begin{equation}\label{eq:leakage}
U_{\bar{1}\bar{0}} = -\frac{i\delta}{2}\int_{-T/2}^t\frac{\Omega_R}{\sqrt{\Omega_R^2+4E_+^2}}e^{-i\int_0^{t'} E_+dt''}dt',
\end{equation}
with $\Omega_R$ and $E_+$ given by Eqs.\ (\ref{eq:omegaR}) and (\ref{eq:energies}) respectively.

\begin{figure}
\includegraphics[width=8cm]{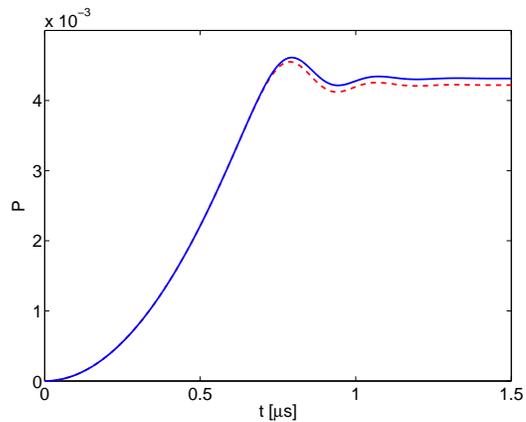}
\caption{(Color online) Population leakage due to $\delta$ calculated perturbatively(dashed line) and by numerical integration(solid line).
Parameters are $\Omega_0 = 4$ MHz, $\Delta=100$ kHz, $\delta=30$ kHz, $\beta = 1.28$ MHz and $\mu = 3$.}\label{fig:pertPop}
\end{figure}
\begin{figure}
\includegraphics[width=7cm]{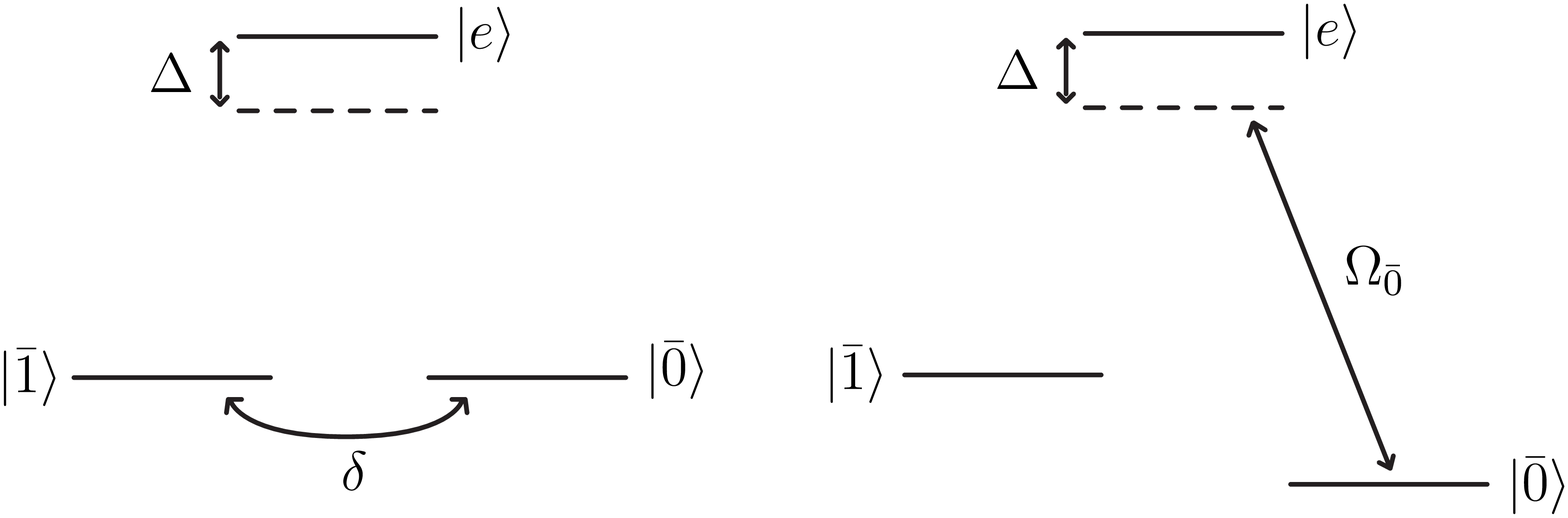}
\caption{When there is no field the weak coupling $\delta$ cannot be ignored, but in the presence of a strong field the coupling of the
hyperfine levels is suppressed by the AC Stark shift.}\label{fig:starkShift}
\end{figure}

The population of the $\ket{\bar{1}}¬£$ state during the pulse is shown in Fig.\ \ref{fig:pertPop}. Note that with the given parameters a naive
estimate of the final population would be $P\sim \delta T = 30$ kHz 1.5 $\mu$s  $= 0.045$ which is an order of magnitude greater than the
observed value $P = 0.0043$. Intuitively we may understand this as an effect of the AC Stark shift. Once the field builds up sufficient strength
the $\ket{\bar{0}}$ level is shifted in energy and the coupling $\delta$ is too weak to drive the transition between non--degenerate states (see
Fig.\ \ref{fig:starkShift}). We thus understand the form of the curve in Fig.\ \ref{fig:pertPop} as follows: at the beginning of the pulse we
observe ordinary Rabi oscillations between the $\ket{\bar{0}}$ and $\ket{\bar{1}}$ states. At $t = 0.59$ $\mu$s we find $E_+ = 10 \delta$ and we
enter the regime $E_+ \gg \delta$ where the curve begins to depart from the ideal sinusoidal shape. At $t = 0.82$ $\mu$s $E_+ = 100 \delta$ and
the curve flattens out and begins to oscillate due to the complex exponential of Eq.\ (\ref{eq:leakage}).

\section{A controlled phase gate}\label{sec:gate}

In the previous section we saw how the AC Stark shift suppresses the dynamics due to $\delta$ during single qubit rotations. Experimental
limitations place an upper limit on the laser power so the suppression cannot be made complete. When working with two--qubit gates we may
restrict ourselves to a single fully entangling gate and thus have a little more freedom in choosing the implementation. In the following we
shall consider a robust implementation of a controlled phase gate.

A naive implementation of a controlled phase gate using the dipole blockade effect is given in Table \ref{tab:naiveGate}.
\begin{table}
\caption{\label{tab:naiveGate}Naive implementation of a controlled phase gate with all pulses implemented as single color Sech pulses. Pulses
2--3 implement an effective $2\pi$--rotation on the target qubit.}
\begin{ruledtabular}
\begin{tabular}{ccccc}
            &  1  &  2  &  3  &  4\\
            \hline
$\varphi_{0,c}$     &  0  &     &     &   $\pi$\\
$\varphi_{1,t}$ &      & 0  &  0 &

\end{tabular}
\end{ruledtabular}
\end{table}
The table is to be read in the following way. First the control qubit is rotated by a Sech pulse through $\pi$ radians about an equatorial axis
of the $\{\ket{e}_c,\ket{0}_c\}$ Bloch sphere. Secondly the target qubit is rotated by two consecutive Sech pulses through $2\pi$ radians about
an equatorial axis of the $\{\ket{e}_t,\ket{1}_t\}$ Bloch sphere and finally the control qubit is rotated through $\pi$ radians retracing its
path on the $\{\ket{e}_c,\ket{0}_c\}$ Bloch sphere. Given an arbitrary initial state
$\ket{\psi}=C_{00}\ket{00}+C_{01}\ket{01}+C_{10}\ket{10}+C_{11}\ket{11}$ the first pulse excites the $\ket{0i}$ components thereby blocking the
subsequent excitation of the $\ket{01}$ component. The second pulse thus only addresses the $\ket{11}$ component, applying a phase of $-1$. The
final pulse returns the $\ket{0i}$ components to their initial state.

\begin{figure}
\includegraphics[width=7cm]{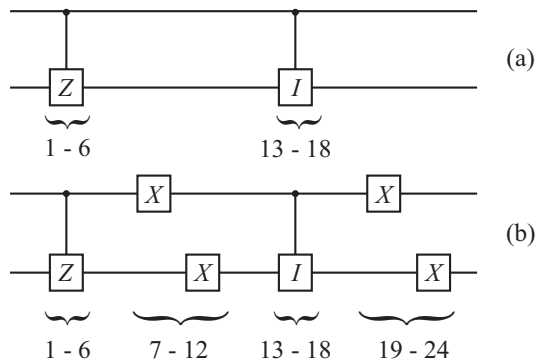}
\caption{(a): Schematic representation of the controlled phase gate implementation of Table \ref{tbl:CPhase} without refocussing pulses. Pulses
1--6 implement a controlled phase gate and pulses 13--18 implement a 'controlled identity' operation which are robust to variations in
$\Delta_t$. Taken together pulses 1--6 and 13--18  are robust to variations in both $\Delta_c$ and $\Delta_t$.\\
(b): Representation of the controlled phase gate sequence with refocussing. Note that since the dipole blockade prohibits simultaneous
excitation of two interacting qubits the NOT gates implemented by pulses 7--12 and 19--24 must be staggered with respect to one
another.}\label{fig:gate}
\end{figure}

Two main sources of errors affect the pulse sequence of Table \ref{tab:naiveGate}. First, the finite channel width causes a detuning dependent
phase on each transition. This defect can be remedied by applying phase compensating pulses \cite{Wesenberg:bus}. Secondly, the hyperfine
broadening causes dephasing of the qubits during gate operation. It is well known from nuclear magnetic resonance studies that such errors can
be suppressed by rapidly flipping the two levels. This approach is not suitable for REQC since we do not have direct access to the $\ket{0}
\leftrightarrow \ket{1}$ transition. Furthermore fast operations are unsuitable since they do not fulfill the adiabaticity criterion. We can,
however, perform a robust bit flip (or NOT gate) using the three--pulse sequence shown in Table \ref{tab:notGate} \cite{Kroll:optCommun}.
\begin{table}
\caption{\label{tab:notGate}Phases $\varphi_0$ and $\varphi_1$ of the single color fields during an implementation of a single qubit NOT gate.}
\begin{ruledtabular}
\begin{tabular}{cccc}
            &  1  &  2  &  3  \\
            \hline
$\varphi_0$ &  0  &     &  0\\
$\varphi_1$ &      &  0  &
\end{tabular}
\end{ruledtabular}
\end{table}
Since the Sech pulse must fulfill the adiabaticity criterion the bit flip is relatively slow. Rather than using the standard 'Bang Bang'
technique we must use the NOT sequence sparingly while retaining some degree of refocussing. To this end we propose the pulse sequence shown in
Table \ref{tab:cPhase}. The sequence consists of a phase compensated controlled phase gate implementation interrupted by a single refocussing
pulse sequence. The effect of the different pulses is summarised in Fig.\ \ref{fig:gate}.
\begin{table}
\caption{\label{tab:cPhase}Implementation of a robust controlled phase gate. Pulses 4 and 5 compensate detuning dependent phase errors caused by
pulses 2 and 3. Pulses 7-12  as well as 19-24 implement a NOT gate on both qubits. Pulses 13-18 compensate detuning dependent errors caused by
pulses 1-6. Note that all pulses are implemented as single color Sech pulses.}
\begin{ruledtabular}\label{tbl:CPhase}
\begin{tabular}{ccccccccccccc}
                            & 1 & 2 & 3 & 4 & 5 & 6 & 7 & 8 & 9 & 10 & 11 & 12\\
                            \hline
$\varphi_{0,c}$ & 0 & & & &  & $\pi$ & 0 &  & 0 &  &  &   \\
$\varphi_{1,c}$ &  &  &  &  &  &  &  & 0 &  &  &  &    \\
$\varphi_{0,t}$  &  & 0 & $\pi$ &  &  &  &  &  &  & 0 &  & 0  \\
$\varphi_{1,t}$  &  &  &  & 0 & 0 &  &  &  &  &  & 0 &    \\
\hline
             & 13  & 14 & 15 & 16 & 17 & 18 & 19  & 20 & 21 & 22 & 23 & 24\\
            \hline
$\varphi_{0,c}$ & 0  &  &  &  & &$\pi$ & 0 & & 0 &  &  &\\
$\varphi_{1,c}$ &  &  &  &  &  &  &  & 0 &  &  &  &\\
$\varphi_{0,t}$  &  & 0 & $\pi$ &  &  &  &  &  &  & 0 &  & 0\\
$\varphi_{1,t}$  &  &  &  & 0 & $\pi$ &  &  &  &  &  & 0 &\\
\end{tabular}
\end{ruledtabular}
\end{table}
In order to evaluate the gate performance we use the fidelity defined by
\begin{equation}\label{eq:fidelity}
\mathcal{F} = \vert \bra{\psi_{in}} U_0^{\dag} U \ket{\psi_{in}}  \vert^2,
\end{equation}
where $\ket{\psi_{in}}$ is the initial state, $U$ the evolution operator implemented by the pulse sequence and $U_0^{\dag}$ the ideal operator
implementing the gate. The fidelity may also be defined as the average of $\mathcal{F}$ or the minimum of $\mathcal{F}$ over the Hilbert space.
We are only interested in a single characteristic of the gate performance, and we shall evaluate Eq.\ (\ref{eq:fidelity}) for a single
superposition input state.

\begin{figure}
\includegraphics[width=8cm]{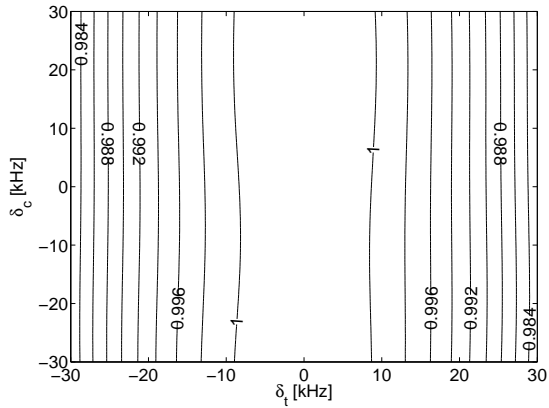}
\caption{Fidelity of the controlled phase gate implementation against the hyperfine shift of the control ($\delta_c$) and target ($\delta_t$)
qubits. The initial state is chosen to be the equally weighted superposition state $\ket{\psi_{in}} =
\tfrac{1}{2}(\ket{00}+\ket{01}+\ket{10}+\ket{11})$ and the pulse parameters are $\Omega_0=4$ MHz, $\beta=1.28$ MHz, $\mu=$ 3, $T=1.5$ $\mu$s,
$\Delta_c=100$ kHz and $\Delta_t=80$ kHz. Note the sequence is virtually insensitive to variations in $\delta_c$.}\label{fig:fid}
\end{figure}

The fidelity of the full gate implementation as a function of the hyperfine shifts of the two qubits is plotted in Fig.\ \ref{fig:fid}. We
observe good fidelities over the relevant parameter range, in particular the fidelity is virtually independent of the shift of the control
qubit. In fact we expect the bit--flip approach to rephasing to work best when the qubit is only mildly perturbed between each flip. Since
almost all non--refocussing manipulations in the gate sequence address the target qubit it is not surprising that it is more vulnerable to the
effects of the hyperfine shift.
If the simulation is run without the refocussing pulses one obtains fidelities over the range $\mathcal{F} = $0.1--1.

\section{Decoherence}\label{sec:decoherence}
For single pulses and short pulse sequences the effects of decoherence can usually safely be ignored. The proposed pulse sequence of Table
\ref{tbl:CPhase} is relatively long and for this reason it is worthwhile to consider decoherence effects in simulations. In this section we
identify the main sources of decoherence in the rare earth quantum computing setup. The hyperfine ground states have very long coherence times,
as long as 82 ms has been reported \cite{Sellars:coherence}. The main source of decoherence is thus attributable to populating the excited
state.

In order to handle the dynamics in the face of relaxations we turn to the density operator formalism and seek to solve the master equation
\begin{equation}\label{eq:master}
\dot{\rho} = i[\rho,H] + \mathcal{L}_{\text{relax}}(\rho).
\end{equation}
The function $\mathcal{L}_{\text{relax}}(\rho)$ is required to be on the so called Lindblad form \cite{Lindblad}
\begin{equation}
\mathcal{L}_{\text{relax}}(\rho) = -\tfrac{1}{2}\sum_m \{C_m^{\dag}C_m \rho + \rho C_m^{\dag} C_m \} + \sum_m C_m \rho C_m^{\dag}.
\end{equation}
The operators $C_m$ describe the relevant decoherence processes. We shall assume that the dominant source of decoherence is spontaneous decay
from the excited state. We thus include the following operators for each ion
\begin{align}
C_0^{(spon)}  &= \sqrt{b_0\Gamma}\vert 0 \rangle \langle e \vert \notag \\
C_1^{(spon)}  &= \sqrt{b_1\Gamma}\vert 1 \rangle \langle e \vert.
\end{align}
The two operators describe spontaneous transitions from the excited state to the two qubit states with branching ratios $b_0$ and $b_1$.

\begin{figure}
\includegraphics[width=8cm]{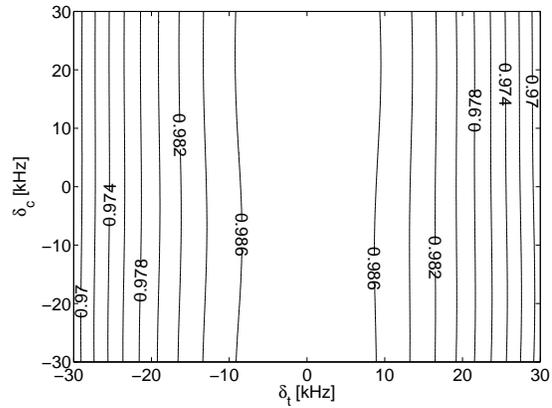}
\caption{Fidelity of the controlled phase gate implementation including decoherence effects. The initial state is chosen to be the equally
weighted superposition state $\ket{\psi_{in}} = \tfrac{1}{2}(\ket{00}+\ket{01}+\ket{10}+\ket{11})$ and the pulse parameters are $\Omega_0=4$
MHz, $\beta=1.28$ MHz, $\mu=$ 3, $T=1.5$ $\mu$s, $\Delta_c=100$ kHz and $\Delta_t=80$ kHz.} \label{fig:fidRelax}
\end{figure}

Finally we solve numerically Eq.\ (\ref{eq:master}) for the full controlled phase gate pulse sequence. A plot of the fidelity as a function of
the hyperfine shifts of the control and target qubits is given in Fig.\ \ref{fig:fidRelax}. We observe a drop in the fidelity of the gate
sequence compared to Fig.\ \ref{fig:fid}. However, we believe we have correctly identified and included the prominent source of decoherence and
so the fidelities of Fig.\ \ref{fig:fidRelax} should closely resemble those obtained in the laboratory. In principle the fidelity could be
further increased by increasing the laser intensity thus allowing faster pulses while remaining in the adiabatic limit. This would be pushing
the limits of current technical feasibility so we do not explore this avenue further.

\begin{figure}
\includegraphics[width=8cm]{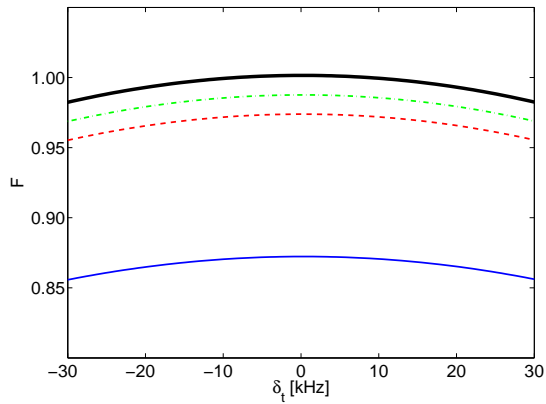}
\caption{(Color online) Fidelity as a function of the target ion hyperfine shift for different excited state lifetimes $T_e$. Starting from
below the solid curve is for $T_e = 100$ $\mu$s, the dashed curve for $T_e = 500$ $\mu$s, the dotted curve for $T_e = 1$ ms and the bold curve
for $T_e = 1$ s.}\label{fig:fidT2}
\end{figure}

It is interesting to analyze the connection between the fidelity and the lifetime $T_e$. In Fig.\ \ref{fig:fidT2} we plot the gate fidelity as a
function of the hyperfine shift of the target ion for different  lifetimes (recall from Figs.\ \ref{fig:fid} and \ref{fig:fidRelax} that the
fidelity is all but independent of the control qubit hyperfine shift). We observe a marked improvement in fidelity as $T_e$ becomes longer, a
point which emphasizes the importance of choosing an ion with the right properties for the experimental realisation.

\section{conclusion}\label{conclusion}

In conclusion we have investigated the effects of random but constant shifts of the hyperfine qubit levels in the REQC system. The evolution of
a two--level system during a complex hyperbolic secant pulse has been analysed analytically to first order using perturbation theory. We have
used these results to explore the consequences of the hyperfine shift during single qubit rotations and have found that errors are dynamically
suppressed due to the AC Stark shift. We have presented a revised implementation of the controlled phase gate which is resilient to the
hyperfine shift and have shown that decent fidelities are possible even when considering the effects of decay and decoherence.

The fidelities are not quite as high as could be desired however this is chiefly due to relaxation and decoherence effects caused by prolonged
integrated population of the excited state which is inevitable if we are to correct for the errors inherent in the system. Recently Wesenberg
\emph{et al.} have proposed a single instance approach to REQC \cite{Wesenberg:scale} which would eliminate some of these errors thereby
allowing higher fidelities.

The gate implementation we have proposed is constructed from intuitive primitives and there is a clear physical idea behind each pulse.
Presumably a more efficient implementation could be found using optimal control theory, possibly using our sequence as a starting point for
numerical optimisation.

\begin{acknowledgments}
We would like to thank the group of Stefan Kr\"oll for valuable discussions concerning technical limitations as well as decoherence and decay
effects inherent in REQC. This work was supported by the European Commission through the Integrated Project FET/QIPC "SCALA".

\end{acknowledgments}

\bibliography{minbib}

\end{document}